\documentclass[lettersize,journal]{IEEEtran}
\usepackage{amsmath,amsfonts}
\usepackage{algorithmic}
\usepackage{algorithm}
\usepackage{color}
\usepackage{array}
\usepackage[caption=false,font=normalsize,labelfont=sf,textfont=sf]{subfig}
\usepackage{textcomp}
\usepackage{booktabs}
\usepackage{stfloats}
\usepackage{url}
\usepackage{verbatim}
\usepackage{graphicx}
\usepackage{cite}
\usepackage{multicol}
\usepackage{multirow}
\usepackage{pifont}
\usepackage{diagbox}
\usepackage{amssymb} 
\begin{document}

\title{MambaDS: Near-Surface Meteorological Field Downscaling with Topography Constrained Selective State Space Modeling}

\author{Zili Liu$^{*}$, Hao Chen$^{\dag}$, Lei Bai, Wenyuan Li, Wanli Ouyang, Zhengxia Zou and Zhenwei Shi$^{\dag}$~\IEEEmembership{Senior Member,~IEEE}
\thanks{The work was supported by the National Natural Science Foundation of China under Grants 62125102, the National Key Research and Development Program of China (Grant No. 2022ZD0160401), the Beijing Natural Science Foundation under Grant JL23005, and the Fundamental Research Funds for the Central Universities, the National Key Research and Development Program of China(Grant No.2022ZD0160101). $^{\dag}$Corresponding author: Zhenwei Shi and Hao Chen (e-mail:shizhenwei@buaa.edu.cn, chenhao1@pjlab.org.cn)}
\thanks{
Zili Liu and Zhenwei Shi are with the Image
Processing Center, School of Astronautics, with the Beijing Key Laboratory
of Digital Media, Beihang University, Beijing 100191, China, and
also with the Shanghai Artificial Intelligence Laboratory, Shanghai 200232,
China.

Hao Chen, Lei Bai, and Wanli Ouyang are with Shanghai Artificial Intelligence Laboratory, Shanghai 200232, China.

Wenyuan Li is with the Department of Geography, University of Hong Kong, Hong Kong, China.

Zhengxia Zou is with the Department of Guidance, Navigation and Control,
School of Astronautics, Beihang University, Beijing 100191, China, and also
with Shanghai Artificial Intelligence Laboratory, Shanghai 200232, China.

$^{*}$This work was done during his internship at Shanghai Artificial Intelligence Laboratory.}}

\markboth{Journal of \LaTeX\ Class Files,~Vol.~14, No.~8, August~2021}%
{Shell \MakeLowercase{\textit{et al.}}: A Sample Article Using IEEEtran.cls for IEEE Journals}


\maketitle

\begin{abstract}
In an era of frequent extreme weather and global warming, obtaining precise, fine-grained near-surface weather forecasts is increasingly essential for human activities. Downscaling (DS), a crucial task in meteorological forecasting, enables the reconstruction of high-resolution meteorological states for target regions from global-scale forecast results. Previous downscaling methods, inspired by CNN and Transformer-based super-resolution models, lacked tailored designs for meteorology and encountered structural limitations. Notably, they failed to efficiently integrate topography, a crucial prior in the downscaling process. In this paper, we address these limitations by pioneering the selective state space model into the meteorological field downscaling and propose a novel model called \emph{MambaDS}. This model enhances the utilization of multivariable correlations and topography information, unique challenges in the downscaling process while retaining the advantages of Mamba in long-range dependency modeling and linear computational complexity. Through extensive experiments in both China mainland and the continental United States (CONUS), we validated that our proposed MambaDS achieves state-of-the-art results in three different types of meteorological field downscaling settings. We will release the code subsequently.
\end{abstract}


\section{Introduction}
\IEEEPARstart{I}{n} recent decades, the increasing frequency and intensity of extreme weather events have underscored the profound impacts of climate change on human societies and natural systems \cite{stott2016climate,lesk2016influence}. Consequently, the need for accurate and reliable weather forecasting has never been more critical. 
Traditional numerical weather prediction methods \cite{bauer2015quiet}, which have been developed over many years, along with the rapidly advancing deep learning-based forecasting models \cite{lam2023learning,bi2023accurate,chen2023fengwu,kurth2023fourcastnet,kochkov2024neural} in recent years, have enabled increasingly accurate global-scale weather predictions. However, due to computational resource constraints, the resolution of global-scale forecasts is limited to tens of kilometers to 100 km \cite{ling2024diffusion}. Such coarse spatial resolution is insufficient for the refined forecasting needs of specific regions \cite{rodrigues2018deepdownscale} and related downstream tasks \cite{younis2008benefit,de2005spatial}.
As a result, using downscaling techniques to generate high-resolution weather forecasts for specific regions is essential for addressing the spatial resolution limitations of global forecast models \cite{wilby1997downscaling,xu2019dynamical,sun2024deep}.

Downscaling methods for meteorological variables can be categorized into dynamical downscaling and statistical downscaling \cite{sun2024deep}. The former, often referred to as Regional Climate Models (RCMs), uses forecast results from Global Climate Models (GCMs) as boundary conditions. It incorporates topography and other regional information to construct regional dynamical processes, solving differential equations to obtain high-resolution meteorological fields at the regional level \cite{pielke2012regional}. However, dynamical downscaling methods heavily rely on differential equations to describe dynamical processes, which often results in significant biases. In addition, the numerical solution process on fine-grained regional grids is computationally intensive, requiring the support of supercomputing platforms \cite{xu2019dynamical}. To alleviate the computational bottlenecks of dynamical downscaling, statistical downscaling methods directly learn the mapping from coarse-resolution to fine-resolution meteorological data. Early attempts use traditional machine learning techniques, such as multiple linear regression \cite{schoof2001downscaling}, random forests \cite{davy2010statistical}, and support vector machines \cite{chen2010downscaling}, to model the mapping relationship. However, the nonlinear mapping capabilities of traditional machine learning methods are relatively limited, which has constrained the performance of early statistical downscaling approaches.

In recent years, machine learning methods, particularly deep learning \cite{lecun2015deep}, have rapidly advanced. Due to their powerful automatic feature extraction capabilities, they have been widely applied in fields such as computer vision \cite{he2016deep}, natural language processing \cite{devlin2018bert}, and earth system science \cite{irrgang2021towards}. The downscaling task of meteorological data draws heavily from the image super-resolution (SR) problem in the field of computer vision. The objective is to recover high-resolution images from low-resolution images, a process very similar to downscaling. Due to the vast amount of meteorological data, which aligns well with the needs of deep learning, deep learning-based image super-resolution methods have seen rapid development and widespread application in the field of meteorological downscaling \cite{sun2024deep}.

Similar to the development of image super-resolution model structures \cite{wang2020deep}, the structures of meteorological downscaling models are also continuously improving \cite{sun2024deep}. Convolutional Neural Networks (CNNs) and their improved methods in SR are the most commonly used model structures for meteorological field downscaling \cite{vandal2017deepsd,hohlein2020comparative,sun2021statistical,tie2022cldassd}. To further enhance the ability to recover fine details and textures, subsequent work has introduced generative adversarial networks (GANs) into the downscaling process \cite{stengel2020adversarial,harris2022generative,hess2022physically}. While CNNs are effective at capturing local nonlinear correlations, their limited receptive field restricts the ability to capture long-term dependencies. Additionally, for large-scale data, it is challenging to continuously improve model performance using CNNs alone. Especially for large-scale meteorological data, the weather conditions of a specific region are often influenced by a broader area. Therefore, the ability to model long-term dependencies is crucial. To enhance long-term dependency modeling and improve fitting capabilities for large-scale data, Transformer-based super-resolution model \cite{liang2021swinir,wang2022uformer} has started to be applied to meteorological downscaling tasks \cite{gerges2022novel, xiang2022spatiotemporal, zhong2024investigating}. However, for high-resolution meteorological fields, the quadratic computational complexity of the self-attention mechanism \cite{vaswani2017attention} in Transformer models leads to substantial memory consumption. Some efficient improvements to the self-attention mechanism can mitigate memory and computational costs to some extent, but they also compromise the model's ability to capture global context \cite{fournier2023practical}. Therefore, designing an efficient model with the ability to capture global context is important for meteorological downscaling tasks.

Recently, state space model (SSM), particularly its improved versions known as Mamba \cite{gu2023mamba}, has garnered significant attention and research in the fields of natural language processing and computer vision due to their powerful global context modeling capabilities and efficient linear computational complexity. \cite{wang2024state, patro2024mamba}. Due to these advantages, SSMs have demonstrated immense potential in high-resolution image processing tasks. Although the original Mamba \cite{gu2023mamba} was designed for modeling one-dimensional (1D) natural language sequence data, subsequent work \cite{zhu2024vision,xu2024survey} has significantly enhanced its capability to process 2D image data by improving the scanning methods within the state space model (SSM). This has allowed it to surpass Transformer-based state-of-the-art (SOTA) methods in several benchmark tasks \cite{wang2024state}. Some early attempts to use the improved Mamba architecture for image SR have also demonstrated that the SSMs could achieve SOTA performance \cite{guo2024mambair,shi2024vmambair,lu2024lfmamba}.
As a result, bringing the Mamba model to the downscaling of meteorological fields to enhance performance is highly significant.

Although downscaling is quite similar to image super-resolution, directly applying SR models to downscaling tasks, as done in previous work \cite{vandal2017deepsd, zhong2024investigating}, has limitations. The fundamental differences between the two primarily lie in the \emph{correlation of multiple variables} and the \emph{influence of auxiliary prior information} \cite{sun2024deep}. The former is mainly because the downscaling process usually involves multiple variables with different distributions (such as wind speed, temperature, precipitation, etc.), and these variables often have certain physical correlations. This is distinctly different from the RGB channels in images. The latter is because fine-scale meteorological states are often influenced by factors such as topography, which can serve as prior information to guide the downscaling process. This is different from super-resolution (SR), where there is typically no obvious prior information that can be utilized.

To address the aforementioned challenges and issues, we introduce \emph{MambaDS}, which pioneers the integration of Mamba to meteorological downscaling, with specialized designs tailored to the unique requirements of downscaling. Specifically, starting from the Visual State Space Module (VSSM) \cite{liu2024vmambavisualstatespace}, we first propose a \emph{Multivariable Correlation-Enhanced VSSM (MCE-VSSM)} by introducing a channel attention branch to establish relationships between different meteorological variables and embed the MCE-VSSM into a SwinIR-like framework \cite{liang2021swinir}, avoiding the texture information loss typically caused by conventional UNet-like structures \cite{cao2022swin, wang2022uformer, shi2024vmambair}. Besides, we design an \emph{efficient topography-constraint layer} at the end of our proposed \emph{MambaDS} to ensure physical consistency, which avoids the extra computational overhead caused by previous methods that integrated terrain information semantically into feature maps \cite{vandal2017deepsd,zhong2024investigating} while ensuring the positive influence of topography priors on downscaling.

To validate the effectiveness of our method, we conducted downscaling experiments with different resolution scales using ERA5 reanalysis data \cite{hersbach2020era5} and NOAA High-Resolution Rapid Refresh (HRRR) analysis/forecast data \cite{dowell2022high} over the regions of mainland China and the continental United States (CONUS). The experimental results confirm the effectiveness of our proposed \emph{MambaDS}, which outperforms previous CNN-based, Transformer-based, and vanilla Mamba super-resolution models in all experimental settings. The main contributions of the paper are as follows:
\begin{itemize}
    \item We propose \emph{MambaDS}, pioneering the integration of the Mamba model into meteorological downscaling, with specialized designs tailored to the unique characteristics of downscaling.
    \item The MCE-VSSM and specially designed topography-constraint layer improve the ability to model multivariable correlations and enhance the efficiency of utilizing topography priors during the downscaling process.
    \item Extensive experiments on ERA5 and HRRR data with different resolution scales demonstrate that the proposed MambaDS achieves state-of-the-art performance in current meteorological field downscaling tasks.
\end{itemize}

\section{Related Work}
This section provides a brief overview of research related to meteorological field downscaling.

\subsection{Deep Models in Meteorological Field Downscaling}

Meteorological field downscaling can be broadly divided into dynamical downscaling and statistical downscaling based on modeling approaches \cite{sun2024deep}. This subsection will focus on the development of deep learning-based methods in statistical downscaling.

In the deep learning modeling paradigm, the downscaling process of meteorological fields is modeled as a nonlinear mapping from low-resolution meteorological grids to high-resolution grids, which is very similar to image SR. Therefore, the development of downscaling models has greatly drawn on the model structures of image SR. CNN-based SR models are the earliest and most common in the application and research of downscaling methods. Representative work includes DeepSD \cite{vandal2017deepsd}, which utilizes the SRCNN \cite{dong2015image} to achieve the mapping from low-resolution to high-resolution meteorological fields. Subsequent improvement methods mostly utilize advanced super-resolution models like VDSR \cite{kim2016accurate} and encoder-decoder-based UNet \cite{ronneberger2015u} to achieve downscaling for meteorological variables such as wind \cite{hohlein2020comparative}, temperature \cite{sha2020deep,doury2023regional}, and precipitation \cite{sha2020deepa,sun2020downscaling}. Additionally, to further enhance the recovery of high-frequency texture information, generative adversarial networks (GANs) \cite{ledig2017photo} have been widely used in downscaling, mainly targeting high-ratio downscaling tasks \cite{stengel2020adversarial} and meteorological variables with richer texture information, such as precipitation \cite{hess2022physically}. Despite these advancements, CNN models have limited capability in modeling global context, which restricts the performance of CNN-based downscaling models.

To further enhance the global context modeling capability of downscaling models, recent work has started using Transformer-based SR models to improve downscaling performance. Zhong utilizes SwinIR \cite{liang2021swinir} and Uformer \cite{wang2022uformer} models to achieve downscaling and bias correction for near-surface temperature and wind speed forecast fields, with model performance significantly surpassing CNN methods \cite{zhong2024investigating}. Other Transformer-based downscaling works have also achieved excellent model performance in areas such as climate variables \cite{gerges2022novel}, spatiotemporal downscaling \cite{xiang2022spatiotemporal}, and station-scale downscaling \cite{liu2024observation, li2024deepphysinet}. However, the quadratic computational complexity caused by the self-attention mechanism in Transformer models requires significant computational power and memory when handling high-resolution images. Some efficient Transformer improvements \cite{fournier2023practical} can alleviate this issue but at the cost of losing the ability to model long-term correlations.

To address the issues of CNN and Transformer models in meteorological field downscaling, we propose \emph{MambaDS}, pioneering the integration of the Mamba \cite{gu2023mamba} into downscaling tasks with special design and improvements. The introduction of the Mamba ensures the ability to maintain long-term dependence while managing model complexity, thereby enhancing the overall performance of the downscaling model.

\subsection{Auxiliary Data for Meteorological Field Downscaling}

Although meteorological field downscaling has borrowed model structures from image SR, the downscaling process is not simply a mapping from low-resolution images to high-resolution images. It is influenced by various physical auxiliary information, such as topography and other meteorological variables \cite{sun2024deep}. Among these, topography information is the most significant factor affecting the meteorological downscaling process. Therefore, it is common in previous work to incorporate it as auxiliary information into the downscaling process. For instance, early precipitation downscaling efforts, such as DeepSD \cite{vandal2017deepsd} and Nest-UNet \cite{sha2020deep}, utilized digital elevation model (DEM) data as auxiliary information. They integrated high-resolution DEM information into feature maps through addition or concatenation, guiding the restoration of detailed texture information. However, such simple integration lacks semantic feature extraction of auxiliary information, limiting the effectiveness of DEM data. Recent improvements \cite{zhong2024investigating} address this issue by designing specialized encoder structures for topography data to extract features at different scales and integrate them into the downscaling process. Although experimental results show that this method can enhance downscaling performance, the additional feature encoder inevitably increases the overall model parameters and computational complexity. This makes it difficult to determine whether the performance improvement comes from increased model capacity or the incorporation of geographical information.

To more efficiently utilize topography information for enhancing and guiding downscaling, we design an \emph{efficient topography constraint layer}. This layer imposes a hard constraint on the downscaled output through topography weighting at the final stage of the model. This approach ensures the effective use of topography information while avoiding additional parameters and computational complexity.

\subsection{Visual State Space Model}

Recently, state space models represented by Mamba \cite{gu2023mamba} have gained significant attention and are widely applied in various fields such as computer vision and natural language processing \cite{wang2024state, patro2024mamba}. Its linear complexity and ability to model long-term dependence help alleviate the inherent issues present in CNN and Transformer models. Although Mamba was initially proposed for sequence data, recent work has demonstrated its strong performance in the visual and image domains \cite{xu2024survey}. Early works of visual Mamba primarily focus on image classification tasks \cite{liu2024vmambavisualstatespace,zhu2024vision,chen2024rsmamba}, addressing the issue of position sensitivity in image information by introducing positional encoding and more complex scanning mechanisms.
Building on this foundation, Mamba has been applied to a wider range of dense prediction tasks, including semantic segmentation \cite{wan2024sigma,zhu2024samba}, remote sensing image change detection \cite{zhao2024rs,zhang2024cdmamba}, medical image segmentation \cite{ma2024u}, and even video tasks \cite{yang2024vivim}. Besides, as an important low-level vision task, Mamba-based image SR works have recently begun to emerge. VMambaIR \cite{shi2024vmambair} embedding an improved SSM module into the UNet structure, achieving state-of-the-art results in image restoration and super-resolution tasks. Similarly, MambaIR \cite{guo2024mambair}, a simple but effective baseline, enhances the performance of vanilla Mamba using local enhancement and channel attention mechanisms. Xiao et al. achieved super-resolution of remote sensing images by introducing an additional frequency selection framework \cite{xiao2024frequency}.

Although Mamba has been widely studied in various visual and image domains, research in the meteorological field is still limited. Particularly for the critical task of meteorological downscaling, it is meaningful to explore how to leverage Mamba's strengths to enhance model performance, considering the specific characteristics of the task.

\section{MambaDS}

This section will provide a detailed introduction to the proposed \emph{MambaDS} structure.

\subsection{Preliminaries: State Space Models}
Original State Space Models (SSMs) \cite{hamilton1994state} are mathematical representations used to describe dynamic systems, extensively applied in control theory, economics, signal processing, and more. They employ a set of first-order differential or difference equations to capture the system's dynamics. A common type of state space model is the Linear Time-Invariant (LTI) system \cite{williams2007linear}, which uses a linear ordinary differential equation (ODE), as shown in Eq. \ref{eq:ssm}, to describe the mapping relationship from the stimulation $x(t)\in\mathbb{R}$ to the response $y(t)\in\mathbb{R}$ through a latent state $h(t)\in\mathbb{R}^N$. 
\begin{equation}
\begin{split}
    h'(t)&={\bf A}h(t)+{\bf B}x(t),\\
    y(t)&={\bf C}h(t)+{\bf D}x(t),
\end{split}\label{eq:ssm}
\end{equation}
where ${\bf A}\in\mathbb{R}^{N\times N}$,${\bf B}\in\mathbb{R}^{N\times 1}$, ${\bf C}\in\mathbb{R}^{1\times N}$, and ${\bf D}\in\mathbb{R}$ represent the parameters of the ODE, $N$ indicates the state size.

To further discretize the continuous differential equation for adaptation to deep learning algorithms, a timescale parameter ${\bf \Delta}$ is used to discretize the continuous parameters ${\bf A}$ and ${\bf B}$ using the Zero-Order Hold (ZOH) criterion. The discretized parameters can be expressed as:
\begin{equation}
    \begin{split}
        {\bf \Bar{A}} &= {\rm exp}({\bf \Delta A}),\\
        {\bf \Bar{B}} &= ({\bf \Delta A})^{-1}({\rm exp}({\bf A})-{\bf I})\cdot {\bf \Delta B}.
    \end{split}
\end{equation}

Therefore, the discretized version of Eq. \ref{eq:ssm} can be expressed as:
\begin{equation}
    \begin{split}
        h_k&={\bf \Bar{A}}h_{k-1}+{\bf \Bar{B}}x_k,\\
        y_k&={\bf C}h_k + {\bf D}x_k,
    \end{split}
\end{equation}
which is consistent with the representation of a recurrent neural network (RNN) \cite{medsker2001recurrent}. However, the inherent structure of RNNs poses challenges for parallel processing. Additionally, the parameters of traditional SSM models are independent of input data, making it difficult to adapt to more complex dynamic systems.

To address the issues mentioned above, recent advancements in state space models, such as S4 and Mamba \cite{gu2023mamba}, have transformed the RNN-based representation into a parallelizable CNN form:
\begin{equation}
    \begin{split}
        {\bf \Bar{K}}&\triangleq ({\bf C\Bar{B}},{\bf C\overline{AB}},\cdots,{\bf C\Bar{A}^{L-1}\Bar{B}}),\\
        {\bf y} &= {\bf x} \ast {\bf \Bar{K}},
    \end{split}
\end{equation}
where ${\bf \Bar{K}}\in\mathbb{R}^{L}$ denotes the structured convolution kernel, $L$ denotes the length of input sequence, $\ast$ is the convolution operation. Additionally, Mamba modifies the parameters ${\bf \Bar{B}}$, ${\bf C}$, and ${\bf \Delta}$ to be input-dependent, enhancing the modeling capability for complex systems. Parallel scan algorithms and hardware-aware algorithms have also been proposed to improve computational efficiency on GPUs.

\subsection{Overall Structure}

\begin{figure*}
\centering
\includegraphics[width=0.9\linewidth]{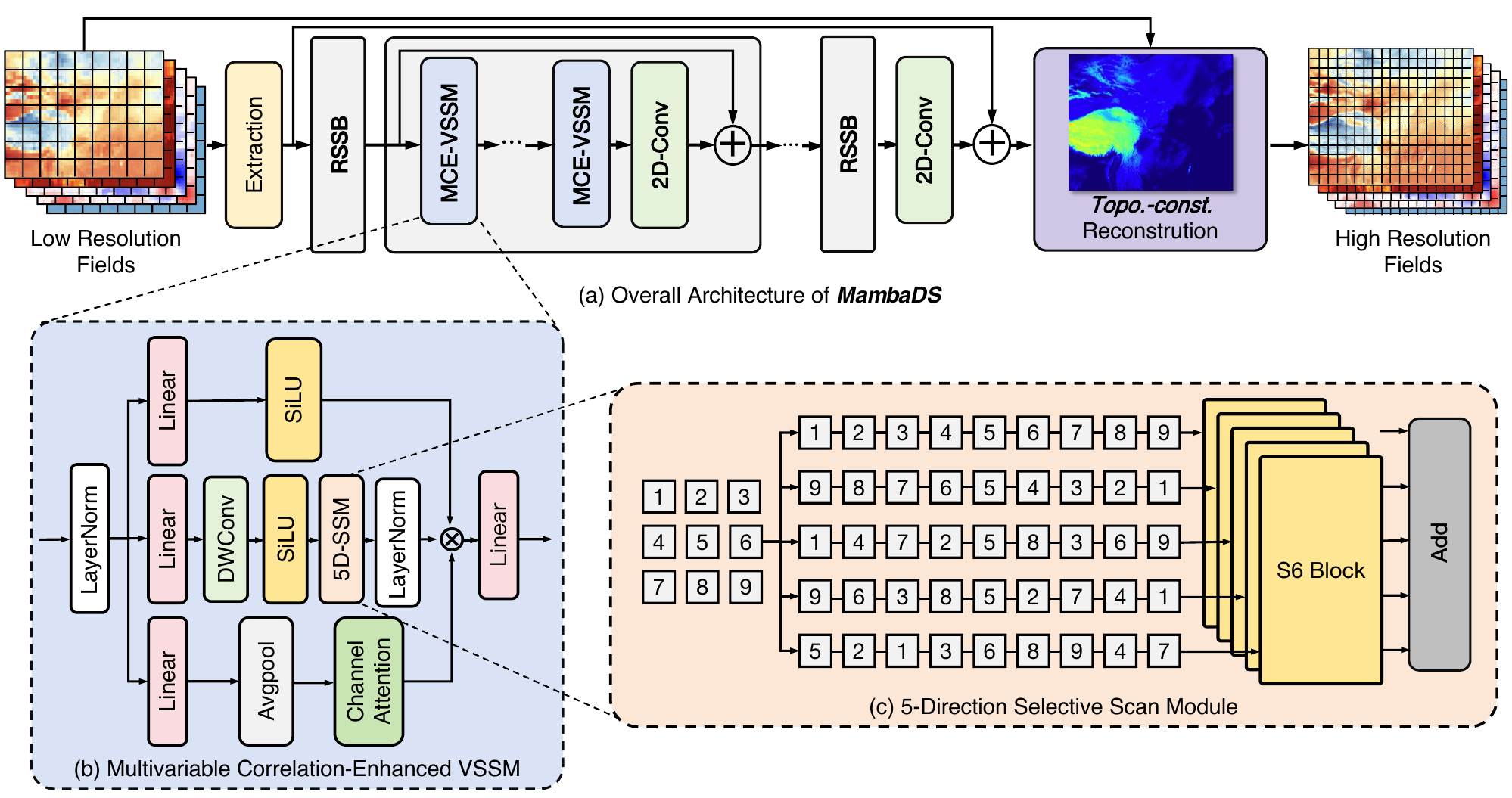}
\caption{Illustration of our proposed \emph{MambaDS} (a), which include three stages. The first shallow feature extraction projects the low-resolution fields into the embedding domain. Then, the stacked Residual State Space Blocks (RSSBs) consisting of multiple Multivariable Correlation-Enhanced VSSMs (MCE-VSSMs) (b) with 5-Direction Selective Scan Module (5D-SSM) (c) are used for deep feature extraction. Finally, the efficient topography-constraint layer is proposed for high-resolution field reconstruction.}
\label{fig:overview}
\end{figure*}

As shown in Fig. \ref{fig:overview}, our proposed \emph{MambaDS} consists of three main stages: a shallow feature extraction, a residual hierarchical Mamba-based encoder, and a topography-constrained reconstruction, which is inspired by the overall architecture of SwinIR \cite{liang2021swinir}. Specifically, given the input low-resolution meteorological fields $\mathcal{F}_{\rm input}\in \mathbb{R}^{h\times w\times V}$, where $h$ and $w$ indicate the number of vertical and horizontal grids, $V$ represents the number of variables. We first utilize a $3\times 3$ convolutional layer to embed the input field to the dimension of $\mathcal{F}_{\rm embed}\in\mathbb{R}^{h\times w\times C}$, where $C$ is the embedding dimension. Subsequently, $\mathcal{F}_{\rm embed}$ extracts deep features through multiple stacked Residual State Space Blocks (RSSBs), each containing several Multivariable Correlation-Enhanced Visual State Space Modules (MCE-VSSMs) and a final convolutional layer to refine extracted features. For each RSSB, the extracted feature is denoted as $\mathcal{F}_{\rm inter}^{l}\in\mathbb{R}^{h\times w\times C}, l\in\{1,2,\cdots,L\}$. The final deep feature obtained could be computed by $\mathcal{F}_{\rm feature}=\mathcal{F}_{\rm inter}^{L}+\mathcal{F}_{\rm embed}$ and is combined with the input low-resolution field $\mathcal{F}_{\rm input}$ to be fed into the topography-constrained reconstruction to achieve a high-resolution meteorological field $\mathcal{F}_{\rm output}\in\mathbb{R}^{H\times\ W\times V}$ with additional topography information.

\subsection{Multivariable Correlation-Enhanced VSS Module}

According to the description above, different meteorological variable fields are treated as channel dimensions of a natural image and stacked together. This approach is also widely used in other AI for meteorology fields, such as weather forecasting \cite{chen2023fengwu}, bias correction \cite{zhong2024investigating}, and assimilation \cite{chen2024fnp}. However, unlike similar spectral channel distributions in images like RGB, different meteorological variables typically exhibit completely different distribution characteristics. Therefore, directly applying image super-resolution methods to multivariable meteorological downscaling tasks, as done in previous downscaling works \cite{zhong2024investigating, vandal2017deepsd}, is limited because it overlooks the different distribution characteristics of each variable. Additionally, there are relationships between variables, often described by atmospheric differential equations.

Therefore, to enhance the modeling capability of correlations between different meteorological variables, as shown in Fig. \ref{fig:overview}(b), we propose a Multivariable Correlation-Enhanced Visual State Space Module (MCE-VSSM). Specifically, given the internal features of the extracted field $\mathcal{F}_{\rm inter}^l\in\mathbb{R}^{h\times w\times C}$, we apply the LayerNorm (${\rm LN}$) \cite{ba2016layer} first and then followed by a three-branched architecture. The output features can be computed by:
\begin{equation}
    \begin{split}
        &\mathcal{\widetilde{F}}_{\rm inter}^l = {\rm LN}(\mathcal{F}_{\rm inter}^l),\\
        &\mathcal{\widetilde{F}}_{\rm inter_1}^l  = {\rm SiLU}({\rm Linear}(\mathcal{\widetilde{F}}_{\rm inter}^l)),\\
        &\mathcal{\widetilde{F}}_{\rm inter_2}^l = {\rm Linear}({\rm Conv}({\rm SiLU}({\rm SSM}({\rm LN}(\mathcal{\widetilde{F}}_{\rm inter}^l))))),\\
        &\mathcal{\widetilde{F}}_{\rm inter_3}^l = {\rm Linear}({\rm Avgpool}({\rm CA}(\mathcal{\widetilde{F}}_{\rm inter}^l)))),\\
        &\mathcal{\widetilde{F}}_{\rm inter_O}^l = {\rm Linear}(\mathcal{\widetilde{F}}_{\rm inter_1}^l\odot \mathcal{\widetilde{F}}_{\rm inter_2}^l\odot \mathcal{\widetilde{F}}_{\rm inter_3}^l),
    \end{split}
\end{equation}
where, ${\rm SSM}(\cdot)$ indicates our proposed 5-Direction Selective Scan Module, ${\rm CA}(\cdot)$ represents the channel attention operation. $\odot$ indicates the Hadamard product. In the above operation, this can be seen as integrating the correlation information between variables into the vanilla VSSM module \cite{liu2024vmambavisualstatespace, shi2024vmambair} in a weighted form through $\mathcal{\widetilde{F}}_{\rm inter_3}^l$, thereby enhancing the model's ability to handle multivariable data.

\subsection{5-Direction Selective Scan Module}

The vanilla Mamba \cite{gu2023mamba}, designed for one-dimensional sequence data in natural language, uses only two directions (forward and backward scanning) to capture the correlation between tokens. However, for image-like 2D grid data with a non-casual nature, both temporal and spatial correlations need to be captured. Therefore, the improved method uses four-directional scanning (top-left to bottom-right, bottom-right to top-left, top-right to bottom-left, and bottom-left to top-right) to enhance the model's spatiotemporal modeling capability \cite{liu2024vmambavisualstatespace}. 

Building on this, to address the chaotic nature of meteorological processes, and inspired by \cite{chen2024rsmamba}, we added a random scanning branch to further enhance the model's ability to capture chaotic dynamic systems (as shown in Fig. \ref{fig:overview}(c)). Specifically, we predefine a random index list, shuffle the input tokens, and fed them into the Mamba module for feature extraction. Then, we restore the original order using the index list and integrate it with other scanning branches.

\subsection{Efficient Topography-Constraint Layer}

\begin{figure*}
\centering
\includegraphics[width=0.8\linewidth]{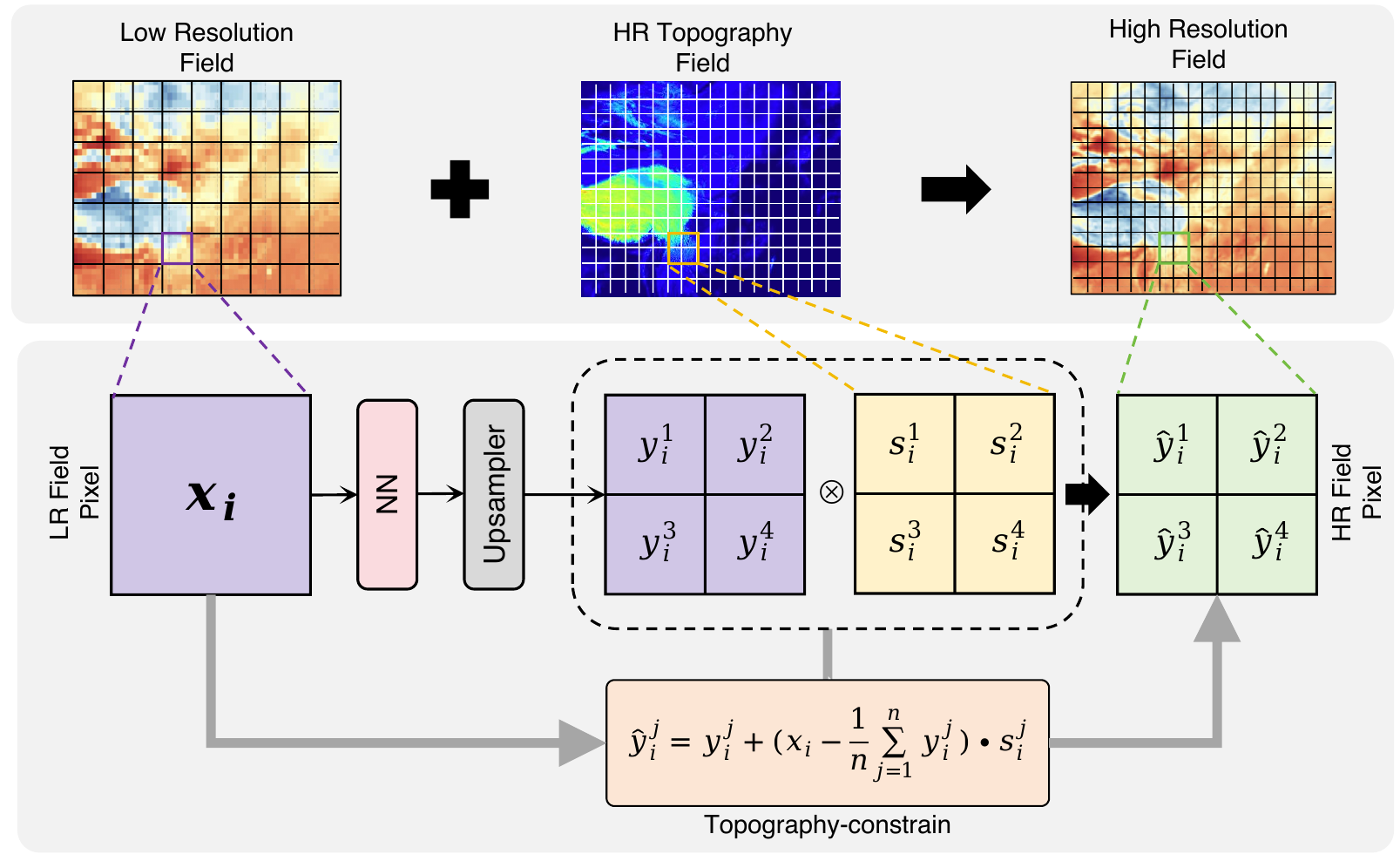}
\caption{Illustration of our proposed efficient topography constraint layer. For each low-resolution pixel $x_i$, predicted high-resolution subpixels $\{y_i^j,j=1,\cdots,n\}$ can be obtained through a downscaling model NN and an upsampler, and then topography constraints on the high-resolution subpixels are achieved using weighted constraints based on topography information.}
\label{fig:constrain}
\end{figure*}

Topography is an important factor and prior information in the downscaling process of meteorological variables. Effectively utilizing high-resolution topographic data to guide the texture restoration of meteorological fields is key to enhancing downscaling performance \cite{sun2024deep}. This characteristic largely reflects the difference between downscaling and general natural image super-resolution, as natural images lack explicit prior texture information that can be utilized. In previous downscaling works \cite{vandal2017deepsd,sha2020deep,zhong2024investigating}, DEM and other topographic data have often been integrated as auxiliary information in the process. However, most studies incorporate topographic information as a separate additional input, directly integrating the raw or independently encoded topographic data into the feature vectors of the hidden layers. Although this approach can implicitly encode geographic information in the feature maps, the spatial information of high-resolution topography is lost during the feature map alignment process. This requires the model to recover detailed textures through encoded semantic features, which increases the learning difficulty to some extent. Additionally, the extra encoding and feature extraction of topography increase the number of parameters and computational complexity of the entire downscaling model.

Therefore, we propose an \emph{efficient topography constraint layer} for better preserving the detailed textures of the topography and avoiding additional computational overhead. Inspired by \cite{harder2023hard}, as shown in Fig. \ref{fig:overview}(a), we placed it directly at the end of the MambaDS model to use topography information to constrain the downscaling model's output. Therefore, the proposed layer can be applied to any other downscaling model in the future to efficiently utilize geographic information and enhance downscaling performance.

Specifically, we use high-resolution DEM data from ETOPO \cite{macferrin2024earth} as topography information and interpolate it to match the spatial resolution of the target meteorological field. As illustrated in Fig. \ref{fig:constrain}, for a given pixel $x_i$ from low-resolution field $\mathcal{F}_{\rm input}$, the downscaling model could produce the corresponding predicted high-resolution pixel $\{y_i^1, y^2_i,\cdots,y_i^n\}$, where $n$ represents the number of high-resolution pixels contained in $x_i$. For example, for 2x downscaling, $n=4$. For $x_i$, we can consider its value as the average of the meteorological states at all positions covered by that pixel, i.e. $x_i = \frac{1}{n}\sum_{j=1}^n y_i^j, n\in\mathbb{N}^+$. Furthermore, topography information can be used as a weighting factor in the averaging process to make it more accurate. Therefore, the general downscaling result constrained by topography information can be calculated as:
\begin{equation}
    \begin{split}
        &\tilde{s}_i^j = {\rm Proj}(s_i^j; \theta), \\
        &\hat{y}_i^j=y_i^j+(x_i-\frac{1}{n}\sum_{j=1}^n y_i^j)\cdot \tilde{s}_i^j,
    \end{split}
\end{equation}
where $s_i^j$ represents the raw pixel at location $(i,j)$ from DEM field $\mathcal{S}\in \mathbb{R}^{H\times W\times 1}$. ${\rm Proj}(\cdot;\theta)$ indicates the projection operation (a CNN layer for implementation) with learnable parameter $\theta$, which provides more flexibility for topography constraints and aligns with the dimension of the number of variables. Through the above operation, the model adjusts the output values based on high-resolution topography information and the mean of coarse-resolution pixels, obtaining a weighted constraint value. This efficiently uses topography information to guide the reconstruction of high-resolution textures.

\section{Results}

This section introduces the study area and dataset used in this paper, as well as the experimental setup and results.

\subsection{Study Area}

\begin{figure*}
\centering
\includegraphics[width=0.9\linewidth]{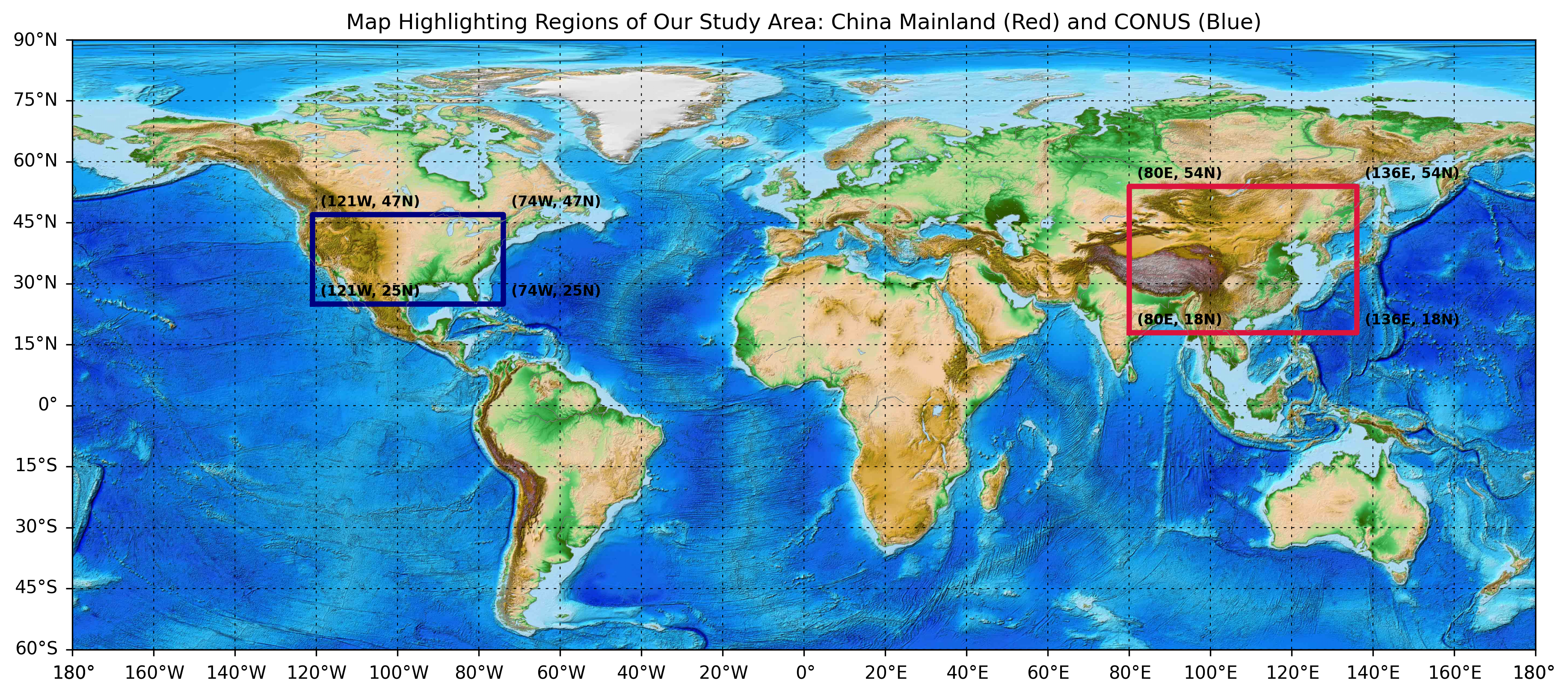}
\caption{The study area in this paper includes two main regions. The red box represents mainland China, with boundaries from $80^{\circ}$E to $136^{\circ}$E and $18^{\circ}$N to $54^{\circ}$N. The blue box represents the CONUS, with longitudes from $74^{\circ}$W to $121^{\circ}$W and latitudes from $25^{\circ}$N to $47^{\circ}$N.}
\label{fig:study_area}
\end{figure*}

As shown in Fig. \ref{fig:study_area}, to fully validate the effectiveness of our proposed method, we selected two regions for our study: China mainland and the Continental United States (CONUS). Specifically, for the mainland China region, we selected the research area with a boundary of $80^{\circ}$E to $136^{\circ}$E and $18^{\circ}$N to $54^{\circ}$N, aligns with the study area of our previous work \cite{liu2024observation}. For the CONUS region, we selected the area with longitudes from $74^{\circ}$W to $121^{\circ}$W and latitudes from $25^{\circ}$N to $47^{\circ}$N. This area is the usable region of the HRRR CONUS data \cite{dowell2022high} in Lambert Conformal projection after geometric correction to a latitude-longitude grid.

We chose two study regions for two main reasons. First, we used different datasets and corresponding resolutions for experiments. These experiments roughly correspond to single-image super-resolution and real-world super-resolution tasks in image SR \cite{guo2024mambair}. This helps to better validate the method's effectiveness. The second reason is that the two regions have significant differences in terrain and location. Testing the downscaling performance in different areas can better demonstrate the method's generalization capabilities.

\subsection{Dataset Description}

In this paper, we use three different datasets to validate our method. Table \ref{tab:dataset} summarizes our dataset.

\begin{table*}[!htbp]
    \centering
    \caption{Dataset and experiment setting in this paper. Dataset period: 2017-01-01 to 2021-12-31.}
    \renewcommand\arraystretch{1.5}
    \resizebox{0.9\linewidth}{!}{
    \begin{tabular}{c|c|c|c|c|c}
    \toprule
         Data Name&Data Type&Resolution&Variables&Experiment Setting&Area  \\
    \hline
         ERA-5 \cite{hersbach2020era5} & Reanalysis & $0.25^{\circ}$ / 1 hour & $u_{10},v_{10},sp,t_{2m},tp_{1h}$ & 4$\times$: $1^{\circ}$ ERA5 $\to$ $0.25^{\circ}$ ERA5 & China\\
         HRRR \cite{dowell2022high} & Analysis & $0.03125^{\circ}$ / 1 hour & $gust, sp, t_{2m}$ & 8$\times$: $0.25^{\circ}$ ERA5 $\to$ $0.03125^{\circ}$ 
 HRRR & CONUS\\
         Fengwu \cite{chen2023fengwu} & Forecast & $0.25^{\circ}$ / 1 hour & $u_{10},v_{10},sp,t_{2m}$ &8$\times$: $0.25^{\circ}$ Fengwu $\to$ $0.03125^{\circ}$ HRRR & CONUS\\
    \bottomrule
    \end{tabular}
    }
    \label{tab:dataset}
\end{table*}

\subsubsection{ERA5 Reanalysis}
ERA5 is a comprehensive reanalysis dataset produced by the European Centre for Medium-Range Weather Forecasts (ECMWF) \cite{hersbach2020era5}. It provides detailed climate data, combining past observations with models to produce consistent and accurate records of atmospheric conditions. ERA5 covers the global climate from 1950 to the present, offering high-resolution data on various parameters such as temperature, precipitation, wind speed, and more. This dataset is widely used for climate research, weather forecasting, and environmental monitoring, offering valuable insights into long-term climate trends and variability.

In this paper, we select data from 5 years and crop the global ERA5 data for the regions of interest and selected 5 commonly used near-surface variables as the focus of the research, which are the wind speed of the u/v component ($u_{10}, v_{10}$), surface pressure ($sp$), air temperature at 2m ($t_{2m}$) and total precipitation in 1 hour ($tp_{1h}$). These five variables are closely related to human activities and industry, and they are also the most widely studied variables in downscaling \cite{sun2024deep}.

\subsubsection{NOAA High-Resolution Rapid Refresh}
The High-Resolution Rapid Refresh (HRRR) is a weather prediction model developed by the National Oceanic and Atmospheric Administration (NOAA) \cite{dowell2022high}. It provides high-frequency, short-term forecasts with a focus on the United States. Updated hourly, HRRR offers detailed data on various meteorological parameters, such as temperature, wind speed, precipitation, and cloud cover, with a high spatial resolution. This model is particularly useful for applications requiring precise, near-term weather predictions, including aviation, severe weather monitoring, and energy management.

In this paper, we use the 0-hour analysis field from the High-Resolution Rapid Refresh (HRRR) model at a resolution of 0.03125 degrees (approximately 3 km) as the downscaling target for the CONUS region. We focus on three near-surface variables: wind speed ($gust$), surface pressure ($sp$), and 2-meter air temperature ($t_{2m}$). It is important to note that the HRRR data does not provide wind speed components or accumulated precipitation data. Therefore, we have excluded these two variables in the CONUS region analysis.

\subsubsection{Fengwu Forecast}
To further validate the effectiveness of the downscaling model in practical applications, we conducted experiments on the downscaling performance for forecast fields. Specifically, we selected the Fengwu medium-range forecast model \cite{chen2023fengwu}, developed by the Shanghai Artificial Intelligence Laboratory, as the generator of the forecast results.

The Fengwu model uses ERA5 reanalysis data as the background field and outputs global 0.25-degree hourly forecasts for the next 10 days through an autoregressive approach. To match the HRRR data, we selected the 6-hour forecast result as input for the downscaling model. Since the Fengwu forecast results are based on ERA5 as the background field, we selected four meteorological variables as input information: $u_{10}$, $v_{10}$, $sp$ and $t_{2m}$. We used the forecast data for 2022 to verify the model's generalization ability for Fengwu forecast results.

\subsection{Experiment Setup}
Building on the proposed MambaDS and the dataset mentioned above, as shown in Table \ref{tab:dataset} column four, we design three types of experiments to evaluate our proposed method: (1) ERA5 reanalysis downscaling, (2) HRRR analysis downscaling, and (3) Fengwu forecast downscaling. In the first experiment, the shape of input and output fields is $5\times36\times 56$ and $5\times 144\times 224$ respectively, and sample label pairs were created by downsampling high-resolution fields. For the second and third experiments, the shape of the input and output fields is $3\times 88\times 188$ and $3\times 704\times 1504$, and the inputs and labels came from different meteorological models. This roughly corresponds to single-image super-resolution and real-world image super-resolution tasks in the image super-resolution field \cite{shi2024vmambair}.

\subsubsection{Architecture Details}
As mentioned in the above section, our proposed MambaDS consists of three stages. During the shallow feature extraction stage, we use a convolutional layer with a kernel size of 3 and a stride of 1 and embed the channel dimension from the number of variables $V$ into $C=240$. Then in the residual hierarchical Mamba-based encoder, we use four RSSBs with depth of $\{14,1,1,1\}$, and the number of channels after each RSSB remains unchanged. Furthermore, the depthwise separable convolution is used in MCE-VSSM to reduce the parameters. Finally, during the reconstruction stage, we upsample the extracted deep features with the pixel shuffle operation together with a convolutional layer. 

\subsubsection{Training Details}
According to the previous works \cite{zhong2024investigating}, we train our proposed MambaDS with the robust Charbonnier loss \cite{charbonnier1994two}:
\begin{equation}
    \mathcal{L} = \sqrt{(y-\hat{y})^2+\epsilon^2},
\end{equation}
where $y$ and $\hat{y}$ represent the predict and ground truth fields, the constant $\epsilon$ is set to $10^{-3}$. This setup allows the loss function to retain the advantages of both L1 and L2 losses. We train the model for 120 epochs and optimized the model using the Adam \cite{kingma2014adam} with $\beta_1=0.9, \beta_2=0.999$, employing a step learning rate initialized as 1e-4 and decreased by half once the number of epochs reaches one of the milestones (60, 96, 108, 114). We trained our proposed model using 4x NVIDIA A100-40G GPUs, setting the batch size to 8 and 2 per GPU on experiments for China mainland and CONUS.

\subsubsection{Evalutate Metrics}
To evaluate the performance of our MambaDS, we follow the previous works \cite{vandal2017deepsd, zhong2024investigating, sun2024deep} and employ four metrics: mean square error (MSE), mean absolute error (MAE), peak signal-to-noise ratio (PSNR) and structural similarity index (SSIM), which are also widely used in image SR. It is important to note that for PSNR and SSIM, we need to adjust the maximum value range of meteorological variables to fit the different value ranges. Therefore, for the China mainland region, we modified the maximum values for different variables to $u_{10}=v_{10}=25m/s, t_{2m}=330K, sp=120000Pa, tp_{1h}=50mm$. For CONUS, $gust=50m/s, t_{2m}=330K, sp=120000Pa$.

\subsection{Performance comparison}
\subsubsection{ERA5 Reanalysis Downscaling}
\begin{table}[!htbp]
    \centering
    \caption{ERA5 reanalysis downscaling results for u/v component wind speed ($u_{10},v_{10}$), surface pressure ($sp$), 2m temperature ($t_{2m}$) and total precipitation in 1 hour ($tp_{1h}$) of various methods.}
    \renewcommand\arraystretch{1.5}
    \resizebox{0.99\linewidth}{!}{
    \begin{tabular}{cc|p{0.06\textwidth}<{\centering}p{0.06\textwidth}<{\centering}p{0.06\textwidth}<{\centering}p{0.08\textwidth}<{\centering}p{0.08\textwidth}<{\centering}}
    \toprule
		 \multicolumn{2}{c|}{{\diagbox{Variable}{Metric}{Method}}} & EDSR \cite{lim2017enhanced} & RCAN \cite{zhang2018image} & SwinIR \cite{liang2021swinir} & VMambaIR$^\star$ \cite{shi2024vmambair} & MambaDS (Ours)\\
    \hline
        \multirow{4}{*}{$u_{10}$}&MSE$\downarrow$&0.1117&0.1004&0.0979&0.1169&{\bf 0.0935}  \\
        &MAE$\downarrow$& 0.2255&0.2145&0.2086&0.2308&{\bf 0.1951}\\
        &PSNR$\uparrow$&37.5357&37.9688&38.1699&37.4885&{\bf 38.5232}\\
        &SSIM$\uparrow$&0.9302&0.9354&0.9385&0.9297&{\bf 0.9432}\\
    \hline
        \multirow{4}{*}{$v_{10}$}&MSE$\downarrow$&0.1170&0.1137&0.1038&0.1208&{\bf 0.0989} \\
        &MAE$\downarrow$ & 0.2314&0.2287&0.2156&0.2411&{\bf 0.2046}\\
        &PSNR$\uparrow$&37.3616&37.5688&37.9736&37.5786&{\bf 38.3097}\\
        &SSIM$\uparrow$&0.9263&0.9289&0.9341&0.9258&{\bf 0.9404}\\
    \hline
        \multirow{4}{*}{$sp$}&MSE$\downarrow$&4813.1685&3824.2467&2508.9448&3424.2675&{\bf 968.9953}  \\
        &MAE$\downarrow$& 51.1968&42.9804&34.5225&39.2877&{\bf 21.8904} \\
        &PSNR$\uparrow$&64.8282&65.8998&67.9306&65.9887&{\bf 72.0241}\\
        &SSIM$\uparrow$&0.9698&0.9745&0.9887&0.9768&{\bf 0.9958}\\
    \hline
        \multirow{4}{*}{$t_{2m}$}&MSE$\downarrow$&0.3710&0.3654&0.3208&0.3775&{\bf 0.2998} \\
        &MAE$\downarrow$& 0.3795&0.3628&0.3443&0.3813&{\bf 0.3126}\\
        &PSNR$\uparrow$&55.0898&55.3287&55.8428&54.2487&{\bf 56.3302}\\
        &SSIM$\uparrow$&0.9772&0.9814&0.9876&0.9759&{\bf 0.9946}\\
    \hline
        \multirow{4}{*}{$tp_{1h}$}&MSE$\downarrow$&0.0624&0.0644&0.0607&0.0622&{\bf 0.0598} \\
        &MAE$\downarrow$ & 0.0544&0.0556&0.0534&0.0542&{\bf 0.0515}\\
        &PSNR$\uparrow$ & 49.6796&49.4785&50.1078&49.6822&{\bf 50.6245}\\
        &SSIM$\uparrow$&0.9848&0.9837&0.9854&0.9843&{\bf 0.9886}\\

    \bottomrule
    \multicolumn{4}{l}{$^\star$ Ours implementation.}
    \end{tabular}}
	\label{tab:era5_metric}
\end{table}

Downscaling the ERA5 reanalysis field is a simple baseline for assessing the performance of downscaling models and has recently been widely used in various meteorological benchmark tasks \cite{ren2023superbench,rasp2020weatherbench}. Although ERA5 data is difficult to obtain in real-time for practical applications, its widespread use and accessibility make experiments based on it meaningful for research and application. Therefore, as shown in Table \ref{tab:dataset}, following our previous work, we selected five important near-surface variables as our research subjects and conducted experiments in the China mainland region. For comparison, we selected four different downscaling models, covering CNN-based \cite{lim2017enhanced,zhang2018image}, Transformer-based \cite{liang2021swinir}, and the recent Mamba-based super-resolution model \cite{shi2024vmambair}. 

\begin{figure*}
    \centering
    \includegraphics[width=1\linewidth]{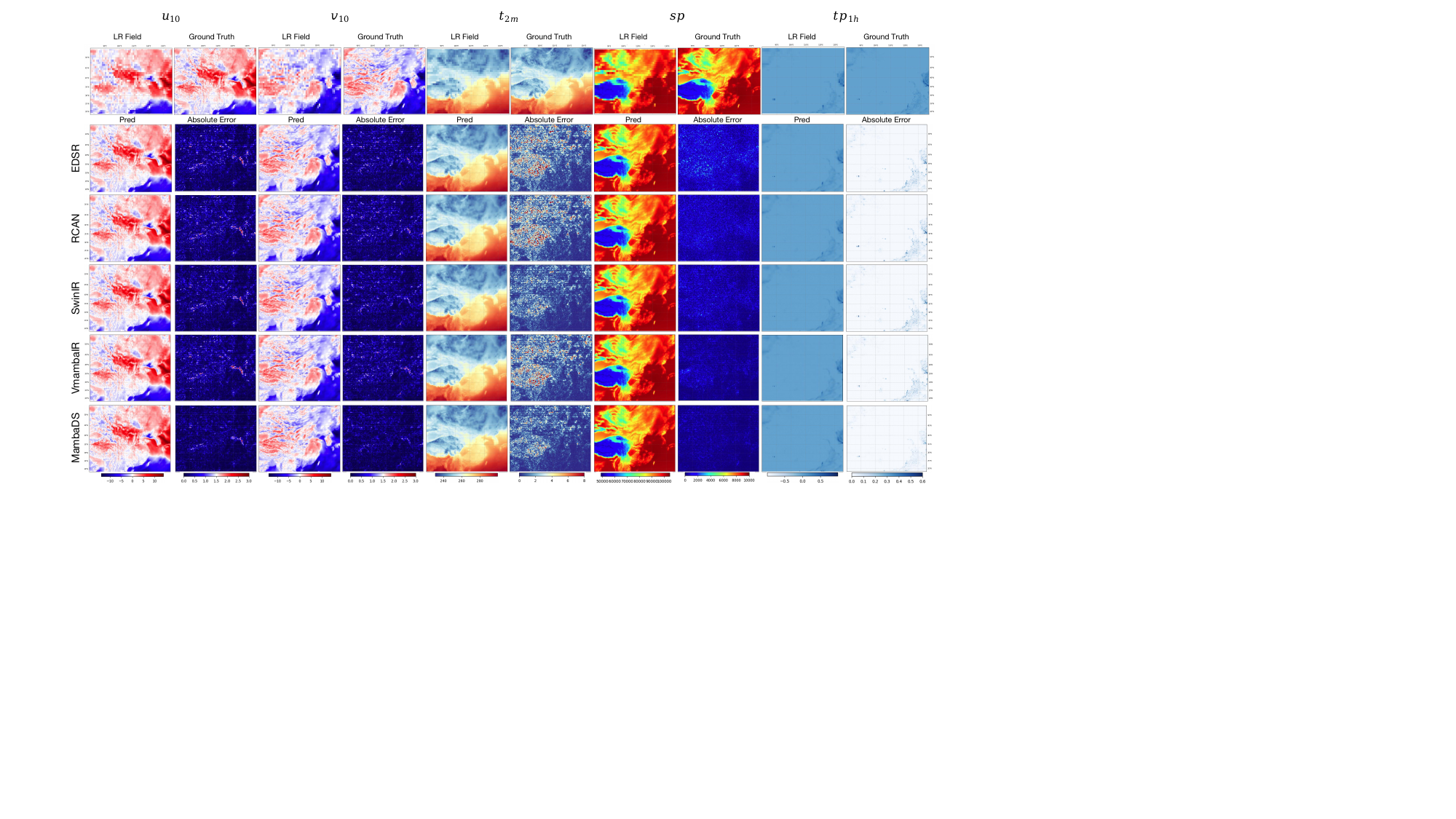}
    \caption{Visualization comparison of ERA5 reanalysis downscaling using different methods. The first row contains the low-resolution meteorological fields of each variable (obtained by downsampling the high-resolution meteorological fields) and the high-resolution GT fields. The following lines show the input results of different downscaling methods and the absolute error map with GT. As can be seen from the figure, the MambaDS proposed in this paper shows the smallest error for all variables, that is, the downscaling performance is optimal.}
    \label{fig:era5}
\end{figure*}

Table \ref{tab:era5_metric} displays the test metrics for five variables using different methods. The results show that the proposed MambaDS method outperforms other methods across all metrics. The results also indicate that different downscaling methods effectively recover known downsampling processes. For instance, with the $t_{2m}$ variable, various models stabilize the MAE around 0.3K. This suggests that the experimental setup is relatively simple. Although different models still show some performance differences, this indicates that even simple tasks can effectively validate the performance of various models. 
It should be noted that another Mamba-based super-resolution model, VMambaIR \cite{shi2024vmambair}, performed worse than the CNN model in this task. We believe that this is because the VMambaIR model is based on the UNet structure, which can improve the efficiency of calculations and reduce the complexity of calculations under the premise of the same number of parameters, but also reduces the model capacity. In addition, the downsampling of feature maps in the UNet structure in the encoder also leads to the loss of detailed texture information. 
Fig. \ref{fig:era5} shows the visualization results of downscaling by different methods and the absolute error with the true value. It can be seen from the figure that the proposed MambaDS performs the best.

\subsubsection{HRRR Analysis Downscaling}

\begin{table}[!htbp]
    \centering
    \caption{HRRR analysis downscaling results for wind speed ($gust$), surface pressure ($sp$), and 2m temperature ($t_{2m}$) of various methods.}
    \renewcommand\arraystretch{1.5}
    \resizebox{0.99\linewidth}{!}{
    \begin{tabular}{cc|p{0.06\textwidth}<{\centering}p{0.06\textwidth}<{\centering}p{0.06\textwidth}<{\centering}p{0.08\textwidth}<{\centering}p{0.08\textwidth}<{\centering}}
    \toprule
		 \multicolumn{2}{c|}{{\diagbox{Variable}{Metric}{Method}}} & EDSR \cite{lim2017enhanced} & RCAN \cite{zhang2018image} & SwinIR \cite{liang2021swinir} & VMambaIR$^\star$ \cite{shi2024vmambair} & MambaDS (Ours)\\
    \hline
        
        \multirow{4}{*}{$gust$}&MSE$\downarrow$&3.4764&3.2453&3.3408&3.5433&{\bf 2.9848} \\
        &MAE$\downarrow$ &1.3534&1.3088&1.3256&1.3457&{\bf 1.059}\\
        &PSNR$\uparrow$&28.8034&29.0973&28.9535&28.8523&{\bf 29.4590}\\
        &SSIM$\uparrow$&0.8596&0.8515&0.8585&0.8588&{\bf 0.8675}\\
    \hline
        \multirow{4}{*}{$sp$}&MSE$\downarrow$&125528.79&150519.64&123167.52&126434.65&{\bf 10474.476}  \\
        &MAE$\downarrow$&171.7839&203.2089&171.3848&172.2387&{\bf 124.3965} \\
        &PSNR$\uparrow$&50.6081&49.8205&50.6904&50.5886&{\bf 50.9918}\\
        &SSIM$\uparrow$&0.9776&0.9763&0.9877&0.9777&{\bf 0.9924}\\
    \hline
        \multirow{4}{*}{$t_{2m}$}&MSE$\downarrow$&3.9440&3.9130&3.7390&3.9385&{\bf 3.4432} \\
        &MAE$\downarrow$&1.3778&1.3715&1.3506&1.3724&{\bf 1.3047}\\
        &PSNR$\uparrow$&44.0257&44.8907& 45.0114&44.2584&{\bf 45.8654}\\
        &SSIM$\uparrow$&0.9817&0.9828&0.9833&0.9824&{\bf 0.9846}\\

    \bottomrule
    \multicolumn{4}{l}{$^\star$ Ours implementation.}
    \end{tabular}}
  
	\label{tab:hrrr_metric}
\end{table}

The downscaling of the HRRR analysis field is a process of learning from the low-resolution ERA5 reanalysis field to the high-resolution analysis field. The two data are obtained from two different modes, so the degradation process of the meteorological field is unknown, which is similar to the real-world image super-resolution task in image super-resolution. It should be noted that since the HRRR analysis field only provides the absolute value of wind speed, but does not provide the component data of wind speed, we merged the wind speed components in ERA5 and obtained the wind speed information at the corresponding location as the input of the model, i.e. $gust=\sqrt{u_{10}^2+v_{10}^2}$. 

Table \ref{tab:hrrr_metric} shows the experimental results of different downscaling methods. From the results, we can see that the error of this experiment is significantly increased compared with the ERA5 downscaling experiment with known degradation processes. For example, for the temperature variable, the MSE increases by about 10 times (from about 0.3K to about 3K). We believe that this is due to the significant differences between the regional assimilation model of HRRR and the global-scale assimilation model of ERA5 in terms of parameterization and discrete grid settings. In addition, as the spatial resolution increases to the kilometer scale, the state of meteorological variables will be more affected by fine-grained scale factors such as topography, thus generating some high-frequency texture information, which is mostly smoothed at the global scale. Although the indicators have changed significantly, the relative performance ranking of different downscaling models remains unchanged. Our proposed MambaDS method still significantly outperforms other methods in all metrics.

\begin{figure*}
    \centering
    \includegraphics[width=1\linewidth]{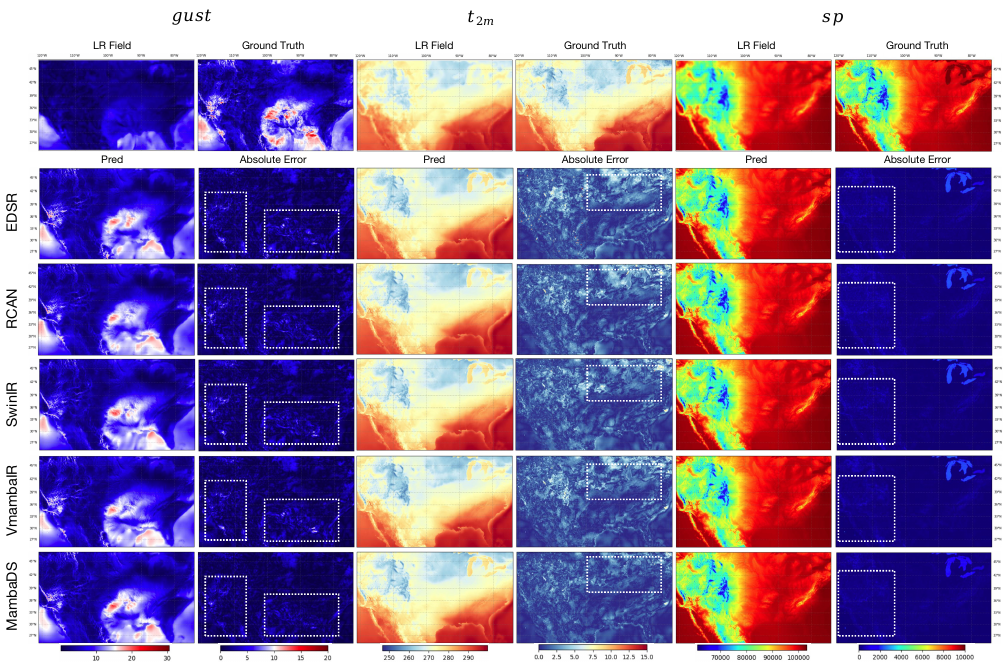}
    \caption{Visualization comparison of HRRR analysis downscaling using different methods. The first row contains the low-resolution meteorological fields of each variable (obtained by downsampling the high-resolution meteorological fields) and the high-resolution GT fields. The following lines show the input results of different downscaling methods and the absolute error map with GT. As can be seen from the figure, the MambaDS proposed in this paper shows the smallest error for all variables, that is, the downscaling performance is optimal.}
    \label{fig:hrrr}
\end{figure*}

Fig. \ref{fig:hrrr} shows the downscaling results of different methods. From the figure, we can see that the output results (LR field and ground truth) of different scale modes are obviously different. This is mainly because the input LR field smoothes most of the high-frequency textures, and the state values at some positions are significantly lower than the HRRR analysis values. Especially for wind speed variables, the ERA5 reanalysis field has obvious underestimates. Therefore, the reconstruction process is more complicated than the downscaling process of ERA5. It can also be seen from the results in the figure that the error distribution range of different variables is closely related to terrain factors. For example, for temperature variables, there is obviously a large error at the junction of land and sea. For pressure, it is more concentrated near the lake (upper right corner). Therefore, it is very important to include terrain data in the downscaling process.

\subsubsection{Fengwu Forecast Downscaling}

\begin{table}[!htbp]
    \centering
    \caption{Fengwu forecast downscaling results for wind speed ($gust$), surface pressure ($sp$), and 2m temperature ($t_{2m}$) of various methods.}
    \renewcommand\arraystretch{1.5}
    \resizebox{0.99\linewidth}{!}{
    \begin{tabular}{cc|p{0.06\textwidth}<{\centering}p{0.06\textwidth}<{\centering}p{0.06\textwidth}<{\centering}p{0.08\textwidth}<{\centering}p{0.08\textwidth}<{\centering}}
    \toprule
		 \multicolumn{2}{c|}{{\diagbox{Variable}{Metric}{Method}}} & EDSR \cite{lim2017enhanced} & RCAN \cite{zhang2018image} & SwinIR \cite{liang2021swinir} & VMambaIR$^\star$ \cite{shi2024vmambair} & MambaDS (Ours)\\
    \hline
        
        \multirow{4}{*}{$gust$}&MSE$\downarrow$&3.9195&3.8374&3.7461&3.8137&\textbf{3.7031}\\
        &MAE$\downarrow$ &1.4411&1.4027&1.4050&1.4298&\textbf{1.3924}\\
        &PSNR$\uparrow$&28.2947&28.6243&28.7245&28.5749&\textbf{28.7425}\\
        &SSIM$\uparrow$&0.8524&0.8544&0.8564&0.8523&\textbf{0.8573}\\
    \hline
        \multirow{4}{*}{$sp$}&MSE$\downarrow$&94139.38&121930.62&91899.40&101245.76&\textbf{80295.97}\\
        &MAE$\downarrow$&162.9230&195.6916&162.2511&163.5641&\textbf{135.3585}\\
        &PSNR$\uparrow$&51.4932&49.2467&50.2414&50.2453&\textbf{52.3823}\\
        &SSIM$\uparrow$&0.9777&0.9734&0.9755&0.9786&\textbf{0.9793}\\
    \hline
        \multirow{4}{*}{$t_{2m}$}&MSE$\downarrow$&5.3664&4.6805&6.1633&6.8214&\textbf{4.5310}\\
        &MAE$\downarrow$&1.7275&1.5707&1.8705&1.9156&\textbf{1.5080} \\
        &PSNR$\uparrow$&42.5323&43.4257&42.1425&42.2465&\textbf{43.6783}\\
        &SSIM$\uparrow$&0.9828&0.9822&0.9824&0.9812&\textbf{0.9884}\\

    \bottomrule
    \multicolumn{4}{l}{$^\star$ Ours implementation.}
    \end{tabular}}
  
	\label{tab:fengwu_metric}
\end{table}

\begin{figure*}
    \centering
    \includegraphics[width=1\linewidth]{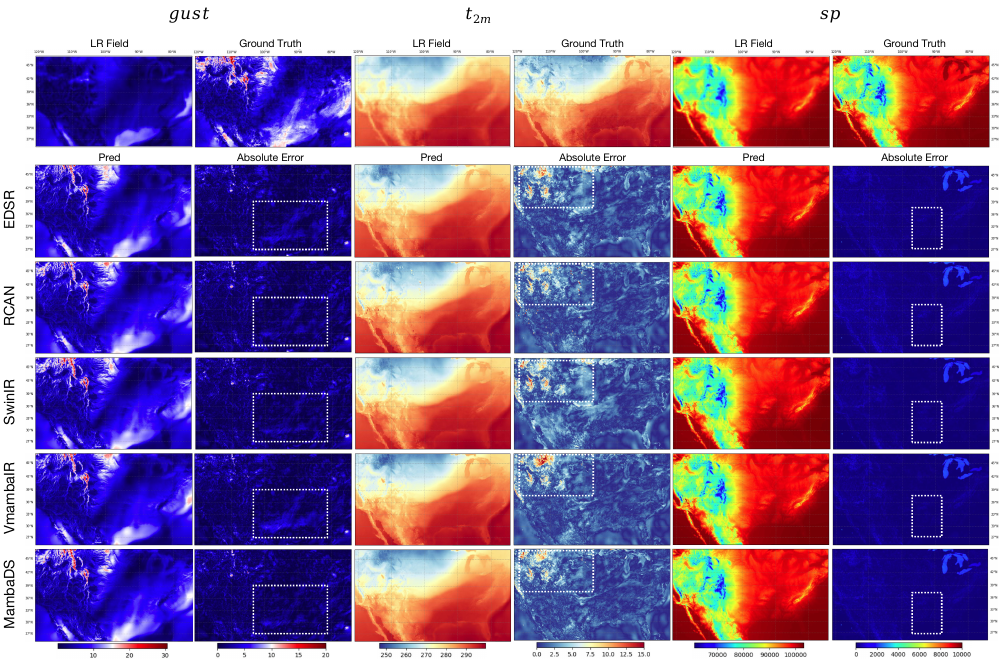}
    \caption{Visualization comparison of Fengwu forecasts downscaling using different methods. The first row contains the low-resolution meteorological fields of each variable (obtained by downsampling the high-resolution meteorological fields) and the high-resolution GT fields. The following lines show the input results of different downscaling methods and the absolute error map with GT. As can be seen from the figure, the MambaDS proposed in this paper shows the smallest error for all variables, that is, the downscaling performance is optimal.}
    \label{fig:fengwu}
\end{figure*}

In order to further verify the effectiveness of the proposed method and get close to the actual downscaling business application scenario, we designed a downscaling task for the forecast field. Instead of using the existing numerical forecast results, we chose Fengwu \cite{chen2023fengwu}, an artificial intelligence medium-term forecast model that has developed rapidly in recent years, as our forecast model. Specifically, we used ERA5 as the forecast initial field of Fengwu and selected the 6-hour forecast results as the input of the downscaling model. It should be noted that we directly use the HRRR analysis field downscaling model to perform zero-shot inference instead of training from scratch, so as to verify the generalization ability of the model to a certain extent. There are three reasons for this setting: First, we believe that the previous two experiments have fully demonstrated the ability of different methods to reconstruct the target input field, and it is redundant to train or fine-tune from scratch based on the Fengwu forecast results. In addition, in actual applications, we usually face forecast results from different models, and their distributions are also different. Therefore, the generalization ability of different downscaling methods for different inputs is more important. Finally, compared with multi-temporal forecast results, the acquisition and storage of the analysis field at time 'f00' is more convenient and economical. A large number of previous works \cite{bi2023accurate,chen2023fengwu} are also mostly based on analysis/reanalysis fields for pre-training to adapt to different downstream inputs.

Table \ref{tab:fengwu_metric} and Fig. \ref{fig:fengwu} show the zero-shot inference results of Fengwu forecast results using different methods. From the results, we can see that the relative performance rankings of different methods for different variables have changed to a certain extent. For example, for the $t_{2m}$, the CNN-based methods EDSR \cite{lim2017enhanced} and RCAN \cite{zhang2018image} outperform the Transformer-based SwinIR \cite{liang2021swinir}, which is different from the results in the first two experiments. Despite this, the MambaDS model proposed in this paper still shows a stable advantage in all variables, especially for $sp$, which is more obvious. We believe that this is also due to the positive effect of adding topography information on surface pressure, a variable that is highly correlated with terrain. The results of this experiment show that the MambaDS proposed in this paper has a stronger generalization ability on the basis of better reconstruction ability than other models. This ability is inseparable from the addition of geographic information and the advantages of the model structure.

\subsection{Ablation Studies}
To further analyze the impact of different components of the proposed MambaDS on the model performance, we conducted ablation experiments. The experimental results listed in Table 3 show that each component in MambaDS improves the performance of the model to varying degrees, and adding all innovative components to MambaDS will achieve the best downscaling performance. Specifically, it can be seen that the addition of the MCE-VSSM module has improved all variables to varying degrees, which proves that compared with the vanilla VSSM module, the enhanced branch of multivariate correlation does provide additional modeling capabilities and model capacity for the downscaling process. As for the addition of the scanning mechanism of 5D-SSM, it can be seen that although it does not show a positive effect on all variables (such as $u_{10}$), it has a positive effect on other variables. This also reflects to a certain extent that this module helps to improve the global modeling capabilities for meteorological downscaling tasks. Finally, the addition of the topography constraint layer can be seen to have the most significant effect on the performance improvement of the model. In particular, for temperature and pressure variables, which are closely related to topography, the addition of terrain information can greatly improve the downscaling performance.

\begin{table*}[!htbp]
    \centering
    \caption{Ablation study results of MambaDS.}
    \renewcommand\arraystretch{1.5}
    \resizebox{0.95\linewidth}{!}{
    \begin{tabular}{c|ccc|cccccccccc}
    \toprule
		 \multirow{2}{*}{Method} & \multicolumn{3}{c|}{Settings}& \multicolumn{2}{c}{$t_{2m}$} & \multicolumn{2}{c}{$sp$} & \multicolumn{2}{c}{$u_{10}$} & \multicolumn{2}{c}{$v_{10}$} & \multicolumn{2}{c}{$tp_{1h}$}\\
   
         &MCE-VSSM&5D-SSM&Topo.& MSE & MAE & MSE & MAE & MSE & MAE & MSE & MAE & MSE & MAE\\
    \hline
        \multirow{4}{*}{MambaDS}&\ding{55}&\ding{55}&\ding{55}&0.3406&0.3552&2137.8452&32.9356&0.1016&0.2173&0.1048&0.2217&0.0612&0.0539\\
   
        & \ding{51}&\ding{55}&\ding{55}&  0.3256 & 0.3462 & 1844.8749 & 28.8756 & {0.1008} & 0.2025 & 0.1044 & 0.2145& 0.0602 & 0.0541\\
    
        & \ding{51}&\ding{51}&\ding{55}&  0.3148 & 0.3414 & 1749.3832 & 27.9723 &0.1012 & 0.2137 & 0.1037 & 0.2128 & 0.0606 & 0.0533 \\
        & \ding{51}&\ding{51}&\ding{51}&  \textbf{0.2998} & \textbf{0.3126} & \textbf{968.9953} &{\bf 0.0935}  & {\bf 0.1951} & \textbf{0.0989} & \textbf{0.2046} & \textbf{21.8904} & \textbf{0.0598} & \textbf{0.0515}  \\
    \bottomrule
    \end{tabular}}\label{tab:ablation_study}
\end{table*}

\subsection{Comparison of different topography prior integration methods}
In the downscaling of meteorological variables, topography data is important prior information. To verify the effectiveness of the topography constraint layer ('soft-Topo.' in Table \ref{tab:topo}) proposed in this paper, we compare our approach with previous topography fusion methods based on feature extraction \cite{zhong2024investigating} ('hard-Topo.' in Table \ref{tab:topo}). Previous feature extraction-based methods use an independent CNN encoder to extract multi-scale topography features and fuse the features of the corresponding scales into the downscaling model. This can be seen as a soft constraint to achieve the constraints on the topography. We compare the downscaling performance of the two methods under the ERA5 downscaling setting and further compare the performance of using the two methods simultaneously ('hard\&soft-Topo.' in Table \ref{tab:topo}).

From the results in Table 3, it can be seen that compared with the model without incorporating topography information ('wo Topop.'), the introduction of terrain information can significantly improve the downscaling effect of different variables. For the a priori integration methods of different approaches, it can be seen that the proposed topography constraint layer achieves downscaling results that are similar to or even exceed those of other methods without a significant increase in the number of parameters. This also further verifies the effectiveness and efficiency of using topography data as a hard constraint weight on the output.

\begin{table}[!htbp]
    \centering
    \caption{ERA5 forecast downscaling results for wind speed ($gust$), surface pressure ($sp$), and 2m temperature ($t_{2m}$) of various topography prior integration methods.}
    \renewcommand\arraystretch{1.5}
    \resizebox{0.99\linewidth}{!}{
    \begin{tabular}{cc|p{0.06\textwidth}<{\centering}p{0.10\textwidth}<{\centering}p{0.10\textwidth}<{\centering}p{0.12\textwidth}<{\centering}}
    \toprule
		 \multicolumn{2}{c|}{{\diagbox{Variable}{Metric}{Method}}} & MambaDS wo Topo. & MambaDS w/ soft-Topo. & MambaDS w/ hard-Topo. (Ours)& MambaDS w/ hard\&soft-Topo.\\
    \hline
        \multicolumn{2}{c|}{Params.(M)} & 14.96&17.04&14.96&17.04\\
    \hline
        \multirow{4}{*}{$u_{10}$}&MSE$\downarrow$&0.1012&0.0997&0.0935&{\bf 0.0933} \\
        &MAE$\downarrow$&0.2137&0.2015&0.1951&{\bf 0.1947}\\
        &PSNR$\uparrow$&38.0154&38.2467&{\bf 38.5232} & 38.5217\\
        &SSIM$\uparrow$&0.9362&0.9402&0.9432&{\bf 0.9433}\\
    \hline
        \multirow{4}{*}{$v_{10}$}&MSE$\downarrow$&0.1037&0.1004&{\bf 0.0989}&0.0993 \\
        &MAE$\downarrow$&0.2128&0.2085&{\bf 0.2046}&0.2050 \\
        &PSNR$\uparrow$&37.9488&38.1958&38.3097&\textbf{38.3846}\\
        &SSIM$\uparrow$&0.9369&0.9396&0.9404&\textbf{0.9411}\\
    \hline
        \multirow{4}{*}{$sp$}&MSE$\downarrow$&1749.3832&1032.8832&\textbf{968.9953}&988.2949 \\
        &MAE$\downarrow$&27.9723&24.5729&\textbf{21.8904}&22.4638\\
        &PSNR$\uparrow$&68.4298&70.2472&72.0241&\textbf{72.3533}\\
        &SSIM$\uparrow$&0.9913&0.9923&0.9958&\textbf{0.9965}\\
    \hline
        \multirow{4}{*}{$t_{2m}$}&MSE$\downarrow$&0.3148&0.3073&0.2998&\textbf{0.2977} \\
        &MAE$\downarrow$& 0.3414&0.3156&0.3126&{\bf 0.3122}\\
        &PSNR$\uparrow$&56.0368&56.2465&56.3302&{\bf 56.3422}\\
        &SSIM$\uparrow$&0.9903&0.9924&0.9946&{\bf 0.9953}\\
    \hline
        \multirow{4}{*}{$tp_{1h}$}&MSE$\downarrow$&0.0606&0.0593&0.0598&\textbf{0.0588} \\
        &MAE$\downarrow$& 0.0533&0.0524&\textbf{0.0515}&0.0517\\
        &PSNR$\uparrow$&49.9935&50.2567&\textbf{50.6245}&50.6105\\
        &SSIM$\uparrow$&0.9842&0.9857&0.9886&\textbf{0.9893}\\

    \bottomrule
    \end{tabular}}
  
	\label{tab:topo}
\end{table}

\section{Conclusion}

In this paper, we pioneer the selective state space model into meteorological field downscaling and propose a novel downscale model namely \emph{MambaDS}.  Compared to previous downscaling methods based on CNN and Transformer super-resolution models, MambaDS can model long-range dependencies while maintaining efficient linear computational complexity. 
Building on this, we made specific designs and improvements tailored to the characteristics of the downscaling task. Specifically, we first designed the MCE-VSSM to enhance the modeling of correlations between different meteorological variables, addressing the need for downscaling multiple variables simultaneously. Additionally, we designed a model-agnostic efficient topography constraint layer that uses hard constraints to directly guide the restoration of fine texture details in meteorological fields with high-resolution topography data, which ensures performance while avoiding additional computational overhead. Through extensive experimental comparisons of three different downscale settings and two study areas, we have verified the effectiveness of our proposed model for multivariable near-surface meteorological field downscaling.

\bibliography{ref.bib}
\bibliographystyle{IEEEtran}

\vfill

\end{document}